# Measuring Data and VoIP Traffic
# in WiMAX Networks

Iwan Adhicandra

**Abstract**—Due to its large coverage area, low cost of deployment and high speed data rates, WiMAX is a promising technology for providing wireless last-mile connectivity. Physical and MAC layer of this technology refer to the IEEE 802.16e standard, which defines 5 different data delivery service classes that can be used in order to satisfy Quality of Service (QoS) requirements of different applications, such as VoIP, videoconference, FTP, Web, etc. The main aim of the paper is to examine a case of QoS deployment over a cellular WiMAX network. In particular, the paper compares the performance obtained using two different QoS configurations differing from the delivery service class used to transport VoIP traffic, i.e. UGS or ertPS. Results indicate that for delay-sensitive traffic that fluctuates beyond its nominal rate, having the possibility to give back some of its reserved bandwidth, ertPS has the advantage to permit the transmission of BE traffic.

———————————— ◆ ————————————

## 1 INTRODUCTION

The IEEE 802.16 technology (WiMAX) is a promising alternative to 3G or wireless LAN for providing last-mile connectivity by radio link due to its large coverage area, low cost of deployment and high speed data rates. The standard specifies the air-interface between a Subscriber Station (SS) and a Base Station (BS). The IEEE 802.16-2004 standard [1], also known as 802.16d, was published in October, 2004. This was further developed into the mobile WiMAX standard referred to as IEEE 802.16e-2005 or 802.16e [2] to support mobile users. IEEE 802.16 can be used not only as a xDSL replacement for small business customers but also as a mobile internet access technology.

One of killer applications for the 802.16 is Voice over IP (VoIP) service to support bidirectional voice conversation. Since its introduction, VoIP has been gaining more and more popularity and some services have been broadened their coverage.

There have been few studies focusing on performance evaluation of IEEE 802.16 WiMAX Networks using OPNET. Ramachandran et al [3] studied performance evaluation of IEEE 802.16 for Broadband Wireless Access. However they used OPNET's DOCSIS models to simulate the IEEE 802.16 MAC. Rangel et al [4] studied performance analysis of QoS scheduling in Broadband IEEE 802.16 Based Networks. Although using OPNET WiMAX models, they focused mainly on implementing their own scheduling algorithms. However, in the IEEE 802.16 standard, the scheduler is left open for implementation, thus creating an avenue for a healthy competition amongst manufacturers. While the standard defines the required procedures and messages for schedulers, it does not offer encouraging means to provide performance, reliability, or

Quality of Service (QoS). Dang et al [5] studied performance of scheduling algorithms for WiMAX networks. Some of their work are quite related with our works. However they focused mainly on implementing some existing scheduling algorithms.

The purpose of this study was to examine a case of QoS deployment over a cellular WiMAX network and to examine the capability of a WiMAX network to deliver adequate QoS to voice and data applications. The methodologies taken include creating the WiMAX network, deploying the required applications, deploying QoS configurations within the WiMAX last-mile, adjusting the QoS configurations within the WiMAX cells to meet voice requirements, and further adjusting the QoS configurations to improve data application performance, without degrading the performance of voice. This topic was identified as being importance to researcher and manufacturers in providing them the necessary background for their works.

The paper is organized as follows. Section 2 explains the QoS in IEEE 802.16. Section 3 provides an overview of VoIP including R-Score and MOS. Section 4 provides the design of system model. The results are presented in Section 5. Finally, Section 6 concludes the paper.

## 2 QUALITY OF SERVICE IN IEEE 802.16

Originally, four different service types were supported in the 802.16 standard: UGS, rtPS, nrtPS and BE. The UGS (Unsolicited Grant Service) is similar to the CBR (Constant Bit Rate) service in ATM, which generates a fixed size burst periodically. This service can be used to replace T1/E1 wired line or a constant rate service. It also can be used to support real time applications such as VoIP or streaming applications. Even though the UGS is simple, it may not be the best choice for the VoIP in that it can waste bandwidth during the off period of voice calls. The rtPS (real-time polling service) is for a variable bit rate

————————————————

- *I. Adhicandra is with the Department of Information Engineering, the University of Pisa, Via Caruso 16, 56127 Pisa, Italy.*

real-time service such as VoIP. Every polling interval, BS polls a mobile and the polled mobile transmits bw-request (bandwidth request) if it has data to transmit. The BS grants the data burst using UL-MAP-IE upon its reception. The nrtPS (non-real-time polling service) is very similar to the rtPS except that it allows contention based polling. The BE (Best Effort) service can be used for applications such as e-mail or FTP, in which there is no strict latency requirement. The allocation mechanism is contention based using the ranging channel. Another service type called ertPS (Extended rtPS) [6] was introduced to support variable rate real-time services such as VoIP and video streaming. It has an advantage over UGS and rtPS for VoIP applications because it carries lower overhead than UGS and rtPS.

## 3 VOICE OVER IP (VoIP) PERFORMANCE

This section will explain the overview of Voice over IP and E-Model.

VoIP application typically works as follows. First, a voice signal is sampled, digitized, and encoded using a given algorithm/coder. The encoded data (called frames) is packetized and transmitted using RTP/UDP/IP. At the receiver's side, data is de-packetized and forwarded to a playout buffer, which smoothes out the delay incurred in the network. Finally, the data is decoded and the voice signal is reconstructed.

The E-model is a transmission planning tool developed by the ITU-T in recommendation G.107 [7]. It provides an expected voice quality prediction as would be perceived by a typical telephone user participating in a complete end-to-end voice call. A wide range of impairments are taken account such as codec impairments, end-to-end delay, jitter, and packet loss, as well as noise and echo.

The ITU-T E-model is based on modeling the results of a large number of subjective tests done to measure perceived voice call quality. It is not a true model in the sense that it cannot accurately predict the absolute opinion of an individual user, but, over a large number of users, the results are sufficiently accurate to permit use for planning and evaluation purposes. The output of the E-model is a value known as the R-value, or Transmission Rating Factor. Other quality measures can also be obtained using this value, such as Mean Opinion Score (MOS). The MOS is a subjective quality score that ranges from 1 (worst) to 5 (best) and is obtained by conducting subjective surveys. The individual transmission parameters are transformed into different impairment factors that are combined to produce the R-value ranging from 0 to 100.

$$R = 100 - Is - Ie - Id + A \qquad (1)$$

where Is is the signal-to-noise impairments associated with typical switched circuit networks paths, Ie is an equipment impairment factor associated with the losses due to the codecs and network, Id represents the impairment caused by the mouth-to-ear delay, and A compensates for the above impairments under various user conditions and is known as the expectation factor.

Once an R-value has been calculated, an estimated MOS can be calculated for the voice call quality using the formulae that follow.

For R < 0, MOS = 1

For 0 < R < 100,

$$MOS = 1+0.035R + 7 \times 10^{-6}R(R - 60)(100 - R) \qquad (2)$$

For R > 100, MOS = 4.5

Variables typically considered in VoIP are only Id and Ie. [8]. So, if default values are used for all other factors, the expression for R-factor from equation (1) can be reduced to:

$$R = 94.2 - Ie - Id \qquad (3)$$

## 4 SYSTEM MODELS

In this experiment, we used OPNET Modeler version 14.5 with WiMAX Module capability [9]. We designed two scenarios including Improve Voice scenario and Improve Data scenario. We assume there are two companies operating the systems. Firstly, a Service Provider company that is a wireless connectivity provider in the region. Secondly, a Client company that has many employees which are mobile throughout a certain area of the region. The Service Provider company needs to assess whether it can accommodate the requirements of the Client company including accessing the corporate servers via a data application and being able to talk to head office via voice application.

*Network Model*

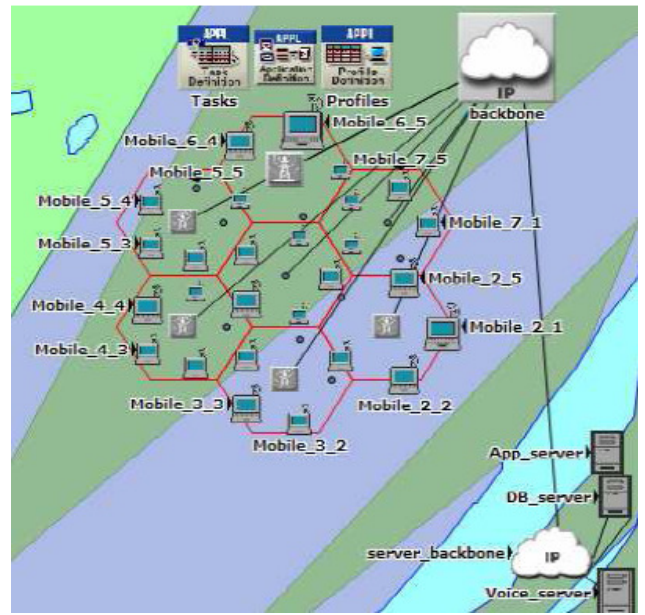

Fig. 1. Network Model.

The nework consists of seven cells and an IP backbone. Cell radius is set to 0.2 km. Each cell has 5 nodes. There is a server backbone containing three servers including Application server, Database server and Voice server. These nodes represent the Service Provider company network (Figure 1). Subscriber Node Transmission Power is set to 0.5 W. Base Station Transmission Power is set to be 5W. The Pathloss and Multipath Model are set to Pedestrian. The parameters of Subscriber Station and Base Station can be seen at Table 1 and Table 2.

<div align="center">

TABLE 1
SS PARAMETERS

</div>

| Attribute | Value |
|---|---|
| ⌐ name | Mobile_1_1 |
| ├ trajectory | NONE |
| ⊟ WiMAX Parameters | |
| ├ Antenna Gain (dBi) | -1 dBi |
| ⊞ Classifier Definitions | (...) |
| ├ MAC Address | Auto Assigned |
| ├ Maximum Transmission Power (W) | 0.5 |
| ├ PHY Profile | WirelessOFDMA 20 MHz |
| ├ PHY Profile Type | OFDM |
| ⊟ SS Parameters | (...) |
| ├ BS MAC Address | Distance Based |
| ⊞ Downlink Service Flows | (...) |
| ⊞ Uplink Service Flows | (...) |
| ├ Multipath Channel Model | ITU Pedestrian A |
| ⊞ Pathloss Parameters | Pedestrian |
| ├ Ranging Power Step (mW) | 0.25 |
| ├ Timers | Default |
| ├ Contention Ranging Retries | 16 |
| ⊞ Mobility Parameters | Default |
| ⊞ HARQ Parameters | (...) |
| ├ Piggyback BW Request | Disabled |
| ├ CQICH Period | 3 |
| └ Contention-Based Reservation Tim... | 16 |

*Application Model*

The Client company will run two applications: a data application consisting of Oracle transactions and a voice application consisting of calls made to the company head office. The Service Provider company has captured the 3-tier Oracle transaction and furnished it to the client company as an ACE trace. The voice traffic is PCM. All client employees will run the data applications, but only one in four will run the voice application in any cell. The data application is deployed between the Client company and the Service Provider company at its head office, using the ACE trace of captured traffic. Then, the PCM-quality voice application is deployed between the Client company and the Service Provider company's head office.

<div align="center">

TABLE 2
BS PARAMETERS

</div>

| Attribute | Value |
|---|---|
| ⌐ name | Base Station_1 |
| ⊟ WiMAX Parameters | |
| ├ Antenna Gain (dBi) | 15 dBi |
| ⊟ BS Parameters | (...) |
| ├ Maximum Number of SS Nodes | 100 |
| ⊞ Received Power Tolerance | (...) |
| ⊞ CDMA Codes | (...) |
| ⊞ Backoff Parameters | (...) |
| ⊞ Mobility Parameters | Default |
| ├ Channel Quality Averaging Parameter | 4/16 |
| ├ Number of Transmitters | SISO |
| ├ ASN Gateway IP Address | Disabled |
| ├ DL AMC Profile Set | Default DL Burst Profile Set |
| ├ UL AMC Profile Set | Default UL Burst Profile Set |
| └ CQICH Period | Accept SS Configured Value |
| ⊞ Classifier Definitions | (...) |
| ├ MAC Address | Auto Assigned |
| ├ Maximum Transmission Power (W) | 0.5 |
| ├ PHY Profile | WirelessOFDMA 20 MHz |
| ├ PHY Profile Type | OFDM |
| └ PermBase | 0 |

<div align="center">

TABLE 3
MAC SERVICE CLASS PARAMETERS

</div>

| Service Class | Type | Maximum Sustained Traffic Rate | Minimum Reserved Traffic Rate |
|---|---|---|---|
| Gold | UGS | 64 Kbps | 64 Kbps |
| Silver | rtPS | 1 Mbps | 0.5 Mbps |
| Bronze | BE | 384 Mbps | 384 Kbps |
| Platinum | UGS | 2.5 Mbps | 2.5 Mbps |

<div align="center">

TABLE 4
CONFIGURED SS PARAMETERS

</div>

| Service Flow | Service Class | Initial Modulation | Initial Coding Rate |
|---|---|---|---|
| DL | Gold | QPSK | 1/2 |
| UL | Gold | QPSK | 1/2 |
| UL | Platinum | QPSK | 1/2 |

*QoS configurations model*

WiMAX QoS in the last mile needs to be configured. We assume that the Service Provider company plans to lease UGS connections with strong QoS guarantees for its voice application. The data application is left to use Best Effort connections without any QoS guarantees. The Client company needs to model their already subscribed bandwidth as a constant UGS bandwidth allocation.

We created a service class Platinum with UGS allocation to reserve the bandwidth the client company has promised their existing employees. They can be accommodated within the remaining bandwidth. The Gold service class is re-sized to the 64Kbps rate of the voice codec (Table 3). Then, we deployed service flows and classifiers on the WiMAX mobile nodes. For each node on which the voice application has been deployed, we deployed UGS connections for both uplink and downlink directions. On Downlink Service Flows and Uplink Service Flow, we set Service Class to Gold, Initial Modulation to QPSK, and Initial Coding Rate to ½. While still in the Uplink Service Flows, we added one extra profile and set Service Class Name to Platinum, Initial Modulation to QPSK, and Initial Coding Rate to ½ (Table 4).

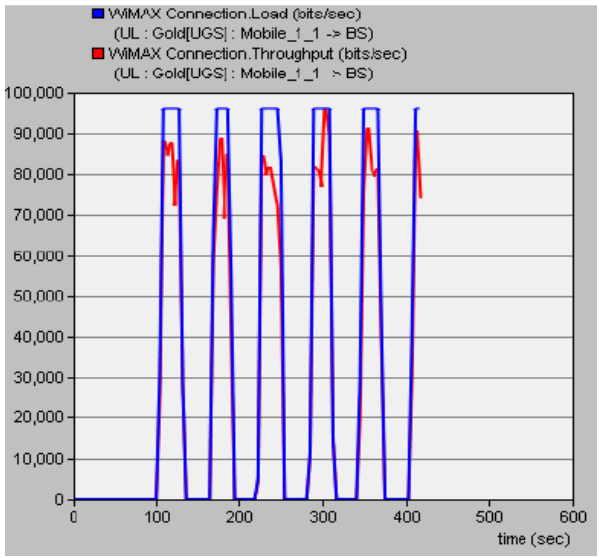

Fig. 2. Voice behaviour over UGS after improvement.

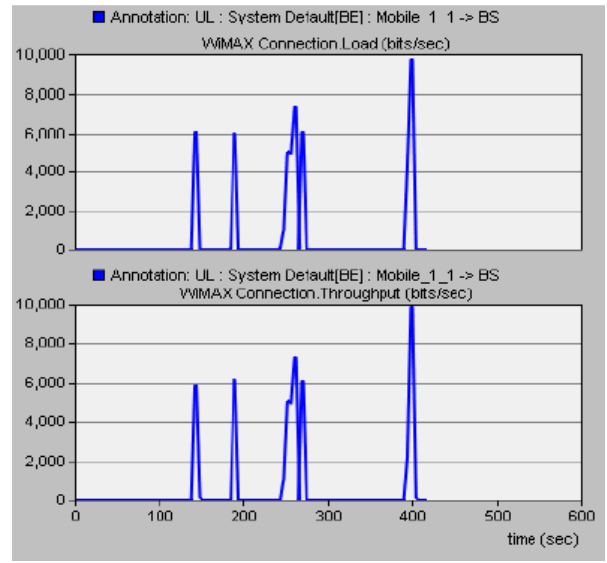

Fig. 4. Data behaviour over Best Effort after improvement.

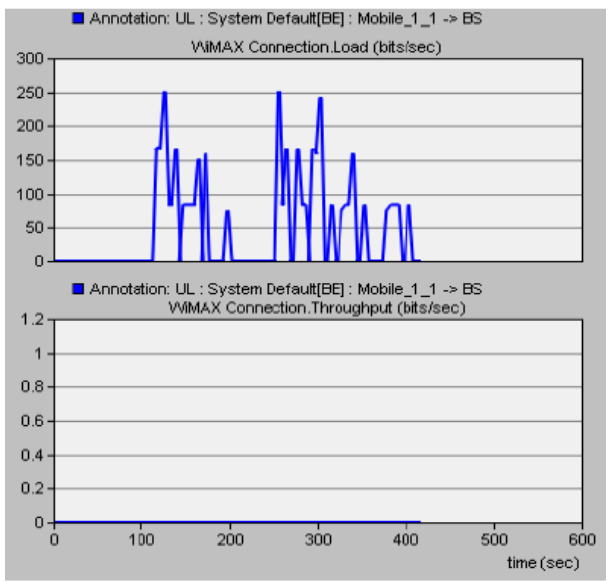

Fig. 3. Behaviour over Best Effort.

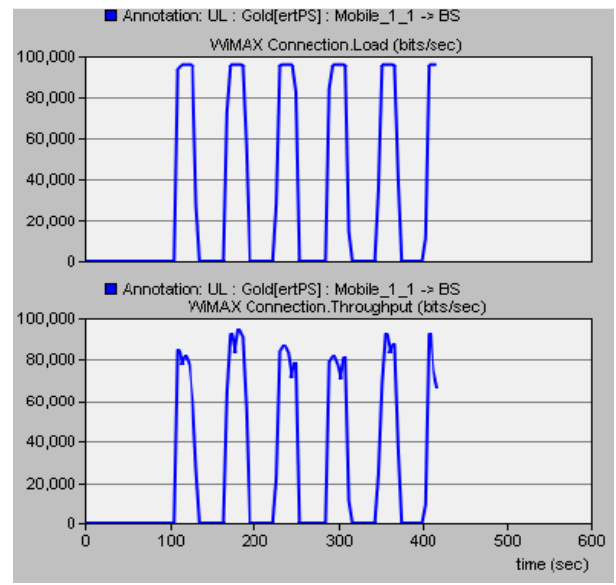

Fig. 5. Voice behaviour over ertPS.

## 5 RESULTS

### 5.1 Improve Voice Scenario

In the normal process, the voice connection has a load of 96 Kbps resulting in the throughput of 64Kbps. This discrepancy between load and throughput is causing unacceptable delays for voice. To solve this problem, we have used the Improve Voice scenario. We re-sized the UGS rate of the Gold service class from 64Kbps to 96Kbps. The throughput is more drawn out in time than the load, suggesting some queuing. The UGS connection was sized for 64Kbps, but it receives 96Kbps in load. The difference is made up by overhead between the application layer and the MAC layer (Figure 2).

Since the voice application offers a peak load at the WiMAX MAC of 96 Kbps, we need to redimension the Gold UGS connection accordingly. Increasing the UGS capacity allows for the voice load at the mobile stations to be matched by the voice throughput at the base station. Some Voice traffic offered as load over the UGS in the MS does not make it through as throughput in the BS. Losses due to interference over the air account for the difference.

Because of the extra capacity invested in the UGS flows carrying voice, the ACE application running in Best Effort stops flowing, even though there is load offered to the Best Effort connection, there is no traffic sent and the connection queue keeps accumulating data packets (Figure 3).

## 5.2 Improve Data Scenario

In the Improve Data scenario, we changed the voice QoS, such that the Best Effort traffic still has a chance to flow. Load is offered to the default Best Effort connection in the uplink. Traffic accumulates in the Best Effort connection queue. No traffic is sent out over the air to relieve the Best Effort queue. This is due to the lack of grants being scheduled for this Best Effort connection. Relief will be provided for the data traffic by changing the scheduling class of the voice connections from UGS to ertPS. The latter provides an elastic allocation, adaptable to the flow of traffic. If there is no traffic load, then the ertPS allocation is reduced temporarily. This reduction provides relief to the Best Effort connection used for the data traffic, without negative impact to the voice application.

In the previous scenario, the Best Effort connection was starved out of grants. Here, as a result of switching voice from UGS to ertPS, traffic started flowing again over the Best Effort connection. Traffic that enters the connection as load on one side emerges as throughput at the other side (Figure 4).

In addition, the voice traffic is not negatively affected (Figure 5). The voice delays in the improve data scenario are only slightly higher than in the improving voice scenario. So, increasing UGS capacity reduces the voice end-to-end delays to below 80ms. Switching from UGS to ertPS does not have too adverse an impact on the voice delays (Figure 6).

For voice quality over different scenarios, we can see that after improving data and voice application, the performance of both voice and data show slightly similar and quite steadily for MOS at around 3.2 until 400 seconds (Figure 7). However, the performance of voice degrade slightly than the performance of data. The reason is because possibly there were some packet losses and jitters during the transactions.

## 7 CONCLUSION

This paper presents our progress to date in evaluating the performance of IEEE 802.16. We have used the WiMAX Connection statistics (e.g. load, throughput) to infer the behavior of traffic mapped to service flows whether the load offered to a connection from the higher layer is matched by the throughput the connection gives back to the higher layer on the other side of the WiMAX hop and whether the connection is starved of grants and traffic from the higher layer builds up in the connection's queue. For delay-sensitive traffic that fluctuates beyond its nominal rate, we have used ertPS scheduling class which has the advantage of giving back some of its reserved bandwidth, if there is no traffic to be served by this bandwidth. For voice quality over different scenarios, the performance of both voice

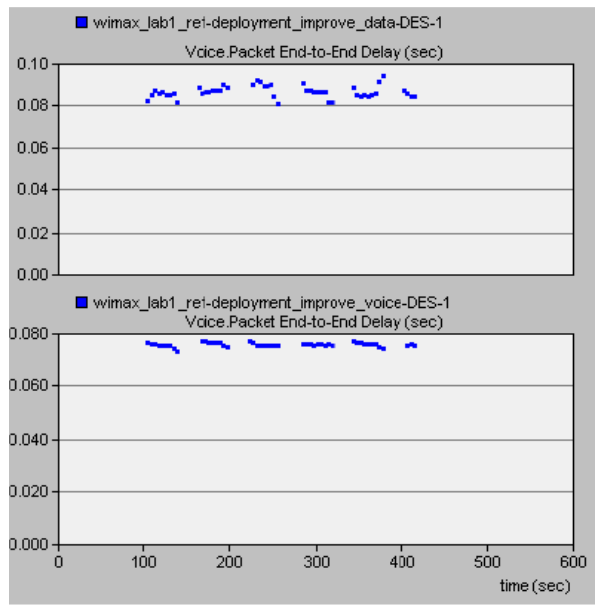

Fig. 6. End-to-end delay for different scenarios.

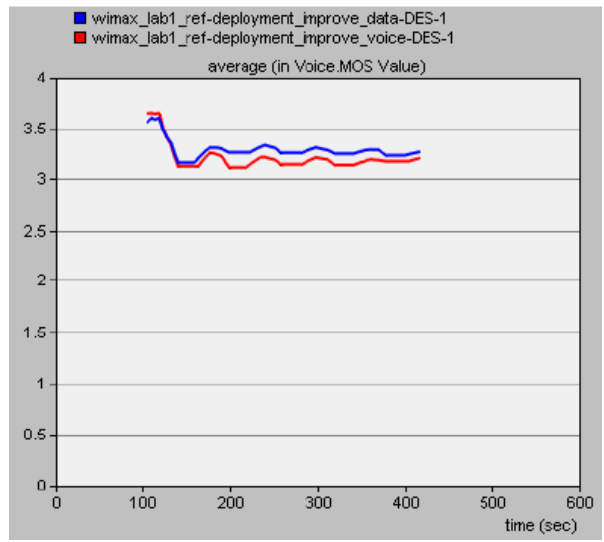

Fig. 7. MOS over different scenarios.

and data show slightly similar and quite steadily for MOS at around 3.2 until 400 seconds.

**Iwan Adhicandra** completed his graduate and postgraduate studies in electrical engineering from Trisakti University, Jakarta, Indonesia, and the University of Sheffield, United Kingdom in 1998 and 2001 respectively. He also completed his postgraduate studies in electrical engineering from Politecnico di Torino, Italy in 2007. From 2000 to 2007 he had been working as a researcher in the Centre for Mobile Communications Research (C4MCR), the University of Sheffield, United Kingdom, in the Computer Communications Research Group, Leeds Metropolitan University, United Kingdom, and in the Connectivity Systems and Networks (COSINE) Group, Philips Research Europe, Eindhoven, the Netherlands. Since October 2007 he has been working with Department of Information Engineering at the University of Pisa, Italy. He is a member of the IEEE, the ACM, the IET, and the BCS.